\documentclass[%
 preprint,
 amsmath,amssymb,
 aps,
prl,
]{revtex4-2}

\usepackage{graphicx}
\usepackage{subcaption}
\usepackage{pgfplots}
\pgfplotsset{compat=1.18}
\usepackage{pgfplotstable}
\usepackage{dcolumn}
\usepackage{bm}
\usepackage[hidelinks]{hyperref}
\usepackage{color} 
\usepackage{siunitx}
\DeclareSIUnit\eangstrom{\text{e\AA}}
\usepackage{amsmath}
\usepackage{braket}
\usepackage{mathtools}
\usepackage[version=4]{mhchem} 
\DeclarePairedDelimiter\abs{\lvert}{\rvert}
\usepackage[normalem]{ulem}
\usepackage[shortlabels]{enumitem}
\usepackage{enumerate}
\usepackage{enumitem}
\usepackage{fixltx2e} 

\newcommand{\bk}{\mathbf{k}}

\newcommand\eqt{\hspace{0.17em}{=}\hspace{0.17em}}

\newcommand{\ezero }{\mathcal{E}_0}
\newcommand{\kres}{\bk_\text{res}}

\begin{document}



\section{Title}

Nonlinear Circular Dichroism Reveals the Local Berry Curvature

\section{Author list}
Nele Tornow$^{1}$, Paul Herrmann$^{1}$, Clemens Schneider$^{2}$, Ferdinand Evers$^{2,3}$, Jan Wilhelm$^{2,\ddagger}$, and {Giancarlo} Soavi$^{1,4,\star}$

\section{Affiliations}
\noindent
$^1$Institute of Solid State Physics, Friedrich Schiller University Jena, Helmholtzweg 5, 07743 Jena, Germany
\newline
$^2$Institute of Theoretical Physics and Regensburg Center for Ultrafast Nanoscopy (RUN), University of Regensburg, Universitätsstrasse 31, 93053 Regensburg, Germany
\newline
$^3$Halle-Berlin-Regensburg Cluster of Excellence CCE, Universitätsstrasse 31,  93053 Regensburg, Germany
\newline
$^4$Abbe Center of Photonics, Friedrich Schiller University Jena, Albert-Einstein-Straße 6, 07745 Jena, Germany
\newline
$^{\ddagger}$ jan.wilhelm@physik.uni-regensburg.de
$^{\star}$ giancarlo.soavi@uni-jena.de

\maketitle


\section{Summary}

Light-matter interactions are governed by conservation laws of energy and momentum. For harmonic generation in crystalline solids, energy conservation imposes that $m$ incoming photons with energy $\hbar \omega_0$ are combined to form one photon at energy $m\hbar \omega_0$. Linear momentum conservation governs phase matching, whereas angular momentum conservation connects the angular momentum carried by photons to the discrete rotational symmetry of the crystal lattice~\cite{Bloembergen.1980, Simon.1968}. As a consequence, circular harmonic generation exerts a torque on the lattice~\cite{Toftul.2023}, and, conversely, a macroscopic rotation of the crystal induces a nonlinear rotational Doppler shift~\cite{Li.2016}. These cornerstone laws of nonlinear optics rely on macroscopic symmetry arguments, and therefore provide little insight into the microscopic origin of angular momentum transfer. Here we uncover a direct connection between angular momentum conservation in nonlinear optics and the electronic quantum geometry, by proving that the transferred angular momentum from light to the crystal is proportional to the local Berry curvature at one optical resonance. This relation is encoded in the nonlinear harmonic circular dichroism, which we measure experimentally in an atomically thin semiconductor. With this, we extend our understanding of nonlinear optics, and we establish a method for the all-optical control and read-out of the local Berry curvature. 

\section{Main text}

The Berry curvature is a manifestation of the gauge invariance of quantum mechanics and is therefore also a central property  of electrons in crystals. In this context it describes quantum geometrical contributions to transport phenomena and optical selection rules, such as the nonlinear Hall effect~\cite{Sodemann.2015, Ma.2019, Kang.2019} and circular dichroism (CD) in time-invariant crystals~\cite{Soavi2025}. While the $\bf{k}$-resolved Berry curvature is straightforward to compute from matrix elements of Bloch wavefunctions~\cite{Xiao2010}, its experimental measurement is challenging, since gauge invariance requires transport observables to involve Brillouin-zone integrals. 
To date, only spin-resolved circular-dichroism angle-resolved photoemission spectroscopy (CD-ARPES) has been proposed as a direct measurement of the $\bk$-local Berry curvature~\cite{Schuler.2020, Beaulieu.2024, Kim.2025, Kang.2025}, even though the range of validity and the resulting limitations of this approach are currently under discussion~\cite{Sidilkover.2025}. A complementary, broadly accessible tabletop, all-optical probe of the Berry curvature with ultrafast time resolution could provide new insights to our understanding of light-matter interactions and quantum geometry. Here, we introduce such an approach via the nonlinear CD in harmonic generation. We show theoretically that for a single resonant crystal momentum the nonlinear CD constitutes a direct probe of the local Berry curvature, which further connects to the angular momentum conservation laws. We furthermore extend the model to the realistic case of multiple $\bk$-resonant optical transitions, and discuss nonlinear CD in relation to the local Berry curvature when time-reversal symmetry (TRS) is broken. 
Finally, we test the predictions of our theory with time-resolved second harmonic generation (SHG) experiments in a prototypical WSe$_2$ monolayer transition-metal dichalcogenide (TMD), where we break TRS by the coherent and valley exclusive optical Stark and Bloch-Siegert effects~\cite{Herrmann2024, Sie.2017}. With these, we measure a value of the Berry curvature of $ \qty{8(2)}{\angstrom\squared}$, in good agreement with theoretical results from tight-binding~\cite{Cho.2018} and density functional theory~\cite{Xiao.2015}, and we thus establish SH-CD as an all-optical, ultrafast, and non-invasive probe of the local Berry curvature.

\subsection{Second harmonic circular dichroism and Berry curvature}
\begin{figure}
    \centering
    \includegraphics[width=90mm]{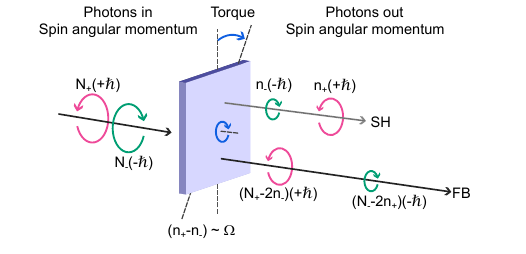}
    \caption{\textbf{Momentum transfer, torque and CD in SHG for a crystal with 3-fold rotational symmetry.} With non-zero SH-CD, a linear FB input generates an elliptical SH output, and an angular momentum $(n_{+}-n_{-}) \cdot 3\hbar$ is transferred from the electromagnetic field to the crystal lattice. Its amplitude $(n_{+}-n_{-})$ is directly proportional to the local Berry curvature. }

    \label{Fig1}
\end{figure}

We work in an independent-particle two-band picture and consider  a single vertical transition at a well-defined crystal momentum
$\bk_{\mathrm{res}}$, which is a good approximation for valley excitons whose momentum-space envelope is
strongly localized around one band extremum, as in monolayer TMDs~\cite{Wang.2018,Man.2021}. We treat $\varepsilon_{cv}(\bk_{\mathrm{res}})$ as an effective
resonance energy given by the exciton transition energy.
At resonance, energy conservation reads $\varepsilon_{cv}(\bk_{\mathrm{res}})
= m\hbar\omega$, and the $m^{th}$ harmonic circular dichroism ($m$H-CD) directly
probes the local Berry curvature (see Supporting Information, Sec.~S1):
\begin{align}
     \text{$m$H-CD} = 
     \frac{|P^+(m\omega)|^2-|P^-(m\omega)|^2}{|P^+(m\omega)|^2+|P^-(m\omega)|^2} =
     \frac{ \Omega(\kres) }{ ||\mathbf{d}_{vc}(\kres)||^2}\,,  \label{e1b}
\end{align}
where $P^{\pm}(m\omega)$ is the circular harmonic polarization, and $\mathbf{d}_{vc}(\kres)$ is the interband dipole matrix element. 
This equation extends the finding of linear optics, $m\eqt 1$~\cite{Yao2008,Cao2012}, and it directly links $m$H-CD to the interband polarization rotation and to the angular momentum conservation law in light-matter interactions. As an exemplary case, we consider SH-CD in a crystal with 3-fold rotational symmetry (Fig.~\ref{Fig1}). In this case, the angular momentum transferred from the electromagnetic field to the crystal lattice equals $(n_{+}-n_{-}) \cdot 3\hbar$, namely an integer multiple of the crystal rotational symmetry. Here, $n_{\pm}$ refers to the number of generated SH photons with left/right circular polarization. On the other hand, $(n_{+}-n_{-}) = \text{SH-CD} \cdot n_{tot}$, where SH-CD is equation~\eqref{e1b} with $m=2$, and $n_{tot}$ is the total number of generated SH photons. This example readily highlights the direct link between local Berry curvature for a single optical resonance and angular momentum conservation in nonlinear optics.

A more realistic situation, involving two energetically degenerate resonances at opposite momenta $\pm \text{K}$, can be experimentally realized in monolayer TMDs, for which we derive a detailed analytical expression of the SH-CD. An off-resonant circularly polarized control beam (CB) lifts the energy degeneracy in the $\pm\text{K}$ valleys and, subsequently, a probe fundamental beam (FB) pulse with photon energy close to half of the optical bandgap generates a resonant SH signal. As discussed, the change in spin angular momentum (SAM) during the nonlinear process is captured by the SH-CD, defined as:

\begin{align}
    \text{SH-CD} = \frac{\abs{\chi^{(2)}_{+ \beta \beta}}^2 - \abs{\chi^{(2)}_{- \beta \beta}}^2}{\abs{\chi^{(2)}_{+ \beta \beta}}^2 + \abs{\chi^{(2)}_{- \beta \beta}}^2} ,
    \label{e2}
\end{align}

where $\chi^{(2)}_{\sigma \beta \beta}$ is the second order nonlinear susceptibility of a $\sigma = \pm$ circularly polarized SH signal, given a linearly polarized FB along the generic Cartesian axis $\beta$. 

\begin{figure}
    \centering
    \includegraphics[width=\linewidth]{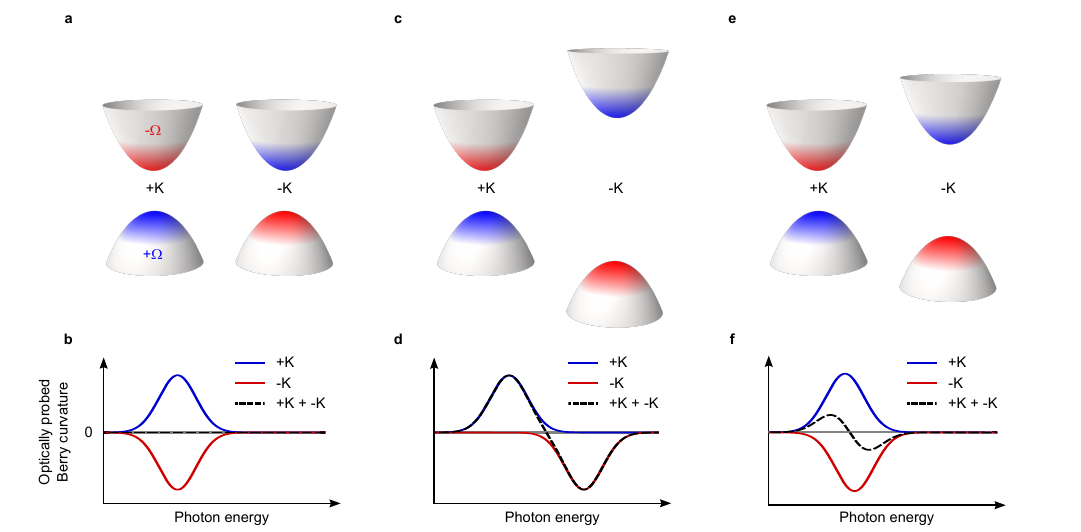}
    \caption{\textbf{Valleys and optically probed Berry curvature in monolayer TMDs.} Equilibrium (\textbf{a}) and TRS broken (\textbf{c,d}) schematic band structure and Berry curvature of a monolayer TMD. Red and blue areas indicate Berry curvature with opposite signs. \textbf{b, d, f,} Optically probed valence-band Berry curvature in each valley (red/blue curves), and their sum (black dashed line) at equilibrium and for broken TRS. The sum of the optically probed Berry curvature at $\pm$K defines the strength of the SH-CD, as captured by equation~\eqref{e2b}.}
    \label{Fig2}
\end{figure}

If the CB is switched off, the energy degeneracy in the $\pm\text{K}$ valleys is preserved (Fig.~\ref{Fig2}~a~and~b), the SH-CD vanishes and a linear input FB generates a linear SH output. Therefore, no momentum is transferred from the electromagnetic field to the crystal, because the optically probed Berry curvature cancels out in $\pm\text{K}$. If the CB is so intense that the two valleys become highly non-degenerate (Fig.~\ref{Fig2}~c~and~d), then the SH-CD is again described by equation~\eqref{e1b}, as for a single optical resonance. For an intermediate situation with a small ($\sim$ 10 meV for typical experimental conditions~\cite{Herrmann2024, Sie.2017}) energy difference between the valleys (Fig.~\ref{Fig2}~e~and~f),  the SH-CD can be calculated analytically by including the valley-dependent optical Stark and Bloch-Siegert shifts into the two-band Hamiltonian of Ref.~\cite{Taghizadeh2019,Herrmann2024}, leading to (see Supporting Information Sec.~S2 for details):

\begin{align}
    \text{SH-CD} &=
   \frac
{\sum\limits_{\bk=\pm \text{K}}w(\bk, \ezero,\omega)\,\Omega (\bk,\ezero) }
{\sum\limits_{\bk=\pm \text{K}}w(\bk,\ezero,\omega)\,||\mathbf{d}_{vc}(\bk,\ezero)||^2 } 
    = \abs{\Omega(\pm \text{K})} \, \sigma \, \ezero^2 \, g(\nu) \, f(\omega) + \mathcal{O}(\ezero^4) \label{e2b}
\end{align}

where $w(\bk,\ezero,\omega)$ is a positive weight that contains the
valley-resolved resonance denominators, $\Omega(\bk,\ezero)$ is the
Berry curvature dressed by the CB field of amplitude $\ezero$, and
$\mathbf{d}_{vc}(\bk,\ezero)$ is the corresponding interband dipole matrix
element. We identify the product $w(\bk, \ezero,\omega)\,\Omega (\bk,\ezero)$ as the optically probed Berry curvature, that we measure in our experiments (panels b, d, f in Fig.~\ref{Fig2}). 

Equation~\eqref{e2b} directly shows the dependence of the SH-CD on the absolute value of the Berry curvature of each unperturbed valley $\Omega(\pm \text{K})$, on the intensity of the pump field ($\sim \ezero^2$), which is a direct consequence of TRS breaking via all-optical coherent effects~\cite{Herrmann2024, Friedrich.2026}, and the sign reversal of the SH-CD upon changing the helicity $\sigma\eqt{\pm}1$ of the CB. The remaining proportionality factors in equation~\eqref{e2b} are $g(\nu)=\nu / (\Delta ^2 (1-\nu^2))$, where $\nu\eqt \omega_\text{CB}/\Delta$  is the normalized CB frequency and $\Delta$ is the equilibrium direct gap, and $f(\omega) = 2+\text{Re}\,\eta(\hbar\omega)$, which describes the detuning of the FB frequency $\omega$ from the equilibrium SH resonance at $\hbar\omega\eqt\Delta/2$ via the complex resonance factor $\eta(\hbar\omega)\eqt {\Delta}/({\Delta{-}2\hbar\omega{-}i\hbar/T_2})$, where $T_2$ is the dephasing time of the optical resonance. 

Considering a FB propagating along the \textbf{z} crystallographic direction and with polarization in the $\textbf{x}$ crystallographic direction,  the SH-CD can be further expressed as:

\begin{align}
    \text{SH-CD} &= 
   \frac
{- 2\text{Im}[\chi^{(2)}_{xxx}(\chi^{(2)}_{yxx})^*] }
{\abs{\chi^{(2)}_{xxx}}^2+\abs{\chi^{(2)}_{yxx}}^2} \, ,
  \label{e4}
\end{align}

indicating that SH-CD can in principle only manifest in the point groups $1$, $3$ and $\overline{6}$ if TRS is preserved, where circular SHG is allowed by momentum conservation and both $\chi^{(2)}_{xxx}$ and $\chi^{(2)}_{yxx}$ are nonzero~\cite{Boyd.2020}. However, it is interesting to note that a further requirement for the SH-CD to be non-zero is that $\chi^{(2)}_{xxx}$ and $\chi^{(2)}_{yxx}$ are not in phase, a property that can't be deduced from a simple macroscopic description of the nonlinear tensors. When TRS is broken, and considering the admissible spin directions~\cite{Friedrich.2026, RodriguezCarvajal.2012}, SH-CD can be non-zero for all non-centrosymmetric magnetic groups with 1- and 3-fold rotational symmetry, i.e. 1, $2'$, $m'$, 3, 3$m'$, 3$2'$, $\overline{6}$, and $\overline{6}m'2'$, where the latter includes the case of monolayer TMDs with broken TRS. For all other magnetic point groups, circular SHG is forbidden either by parity (centrosymmetric crystals), or by angular momentum conservation. Equation~\eqref{e4}, combined with equation~\eqref{e2b}, indicates that an asymmetry between energy levels ($E_{\uparrow} (+\bk) \neq E_{\downarrow} (-\bk)$ and $E_{\uparrow} (+\bk) \neq E_{\uparrow} (-\bk)$) and Berry curvature ($\Omega (+\bk) \neq \Omega (-\bk)$ and $\Omega (+\bk) \neq -\Omega (-\bk)$) at opposite momenta in reciprocal space induces a phase mismatch between the emerging $\chi^{(2)}_{xxx}$ and $\chi^{(2)}_{yxx}$ components of the second order susceptibility. This provides physical insights that link macroscopic symmetries to microscopic features of the electronic band structure and quantum geometry. 

\subsection{Ultrafast all-optical control of nonlinear circular dichroism} 
\begin{figure}[ht]
    \centering
    \includegraphics[width=\linewidth]{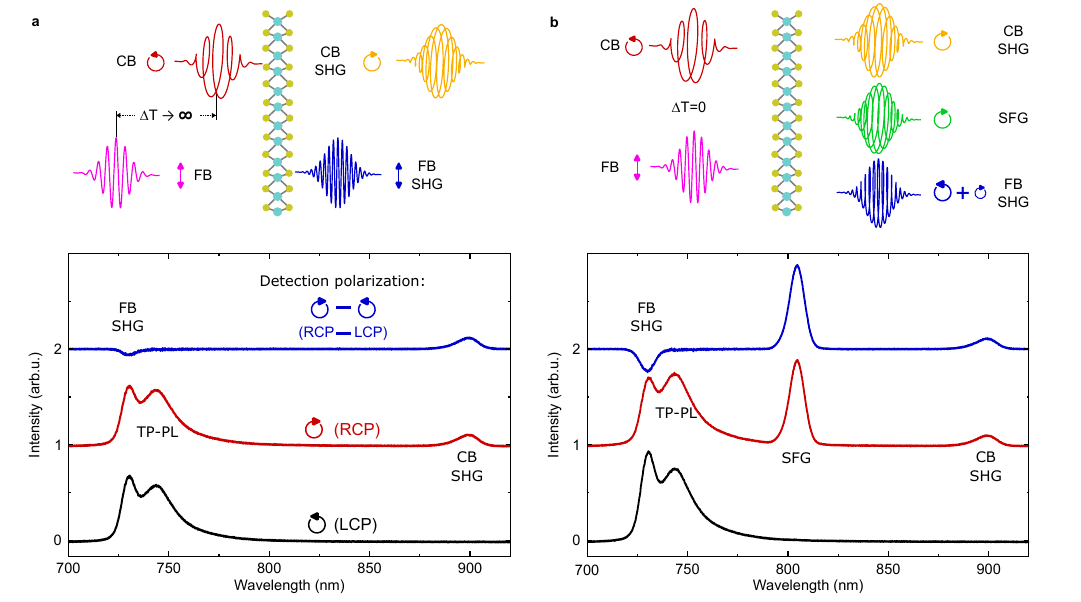}
    \caption{\textbf{Helicity and time-resolved SH spectra.} \textbf{a}, SH spectra at large negative delays between the LCP CB and a linearly polarized FB. The LCP (black curve), RCP (red curve) detected spectra, and their difference (blue curve), show unpolarized TP-PL, the RCP SH signal of the CB and an almost linearly polarized SH signal of the FB. \textbf{b}, Same as in panel (a), but at zero time delay between the CB and the FB. In this case, the SH signal of the FB is clearly elliptical.}
    \label{Fig3}
\end{figure}

We demonstrate ultrafast all-optical control of the SH-CD in a two-color pump-probe experiment in a WSe$_2$ monolayer. Exemplary helicity resolved emission spectra are shown in {Fig.~\ref{Fig3}} for zero (panel b) and long (i.e., much longer than the pulse duration, panel a) delays between a left circularly polarized (LCP) CB and a linearly polarized FB, responsible for SH generation. The CB/FB  wavelengths are fixed at \SI{1800}{\nm}/\SI{1460}{\nm} (see Methods for details). In Fig.~\ref{Fig3}, the black/red curves represent spectra with co-/cross-polarization with respect to the circular CB, while the blue curves show their difference. 
In all cross-polarized spectra, regardless of the CB to FB time delay, we observe the SH of the CB at \SI{900}{\nm}, as expected from SAM conservation~\cite{Bloembergen.1980}. This signal is absent in the co-polarized spectra. At zero delay (Fig.~\ref{Fig3}~b), we further observe a sum frequency generation (SFG) signal between the CB and FB pulses, which is again cross-polarized with respect to the CB.
In the region between approx. \SIrange[]{730}{750}{\nm} we observe the most relevant signal for our study of SH-CD, namely the SH of the FB. This overlaps spectrally with the two-photon photoluminescence (TP-PL), mainly originating from the FB.  When taking the difference between LCP and RCP detection (blue curves in Fig.~\ref{Fig3}), the TP-PL always cancels out due to its incoherent nature at room temperature~\cite{Herrmann.2023}. In contrast, the FB-SH signal appears clearly in the LCP-RCP spectrum at zero delay (panel b), underlying the SH-CD. The signal is drastically reduced at large delays (panel a), where the small residual is most likely due to an initial and uncompensated helicity of the FB.

\begin{figure}[t!]
    \centering
    \includegraphics[width=\linewidth]{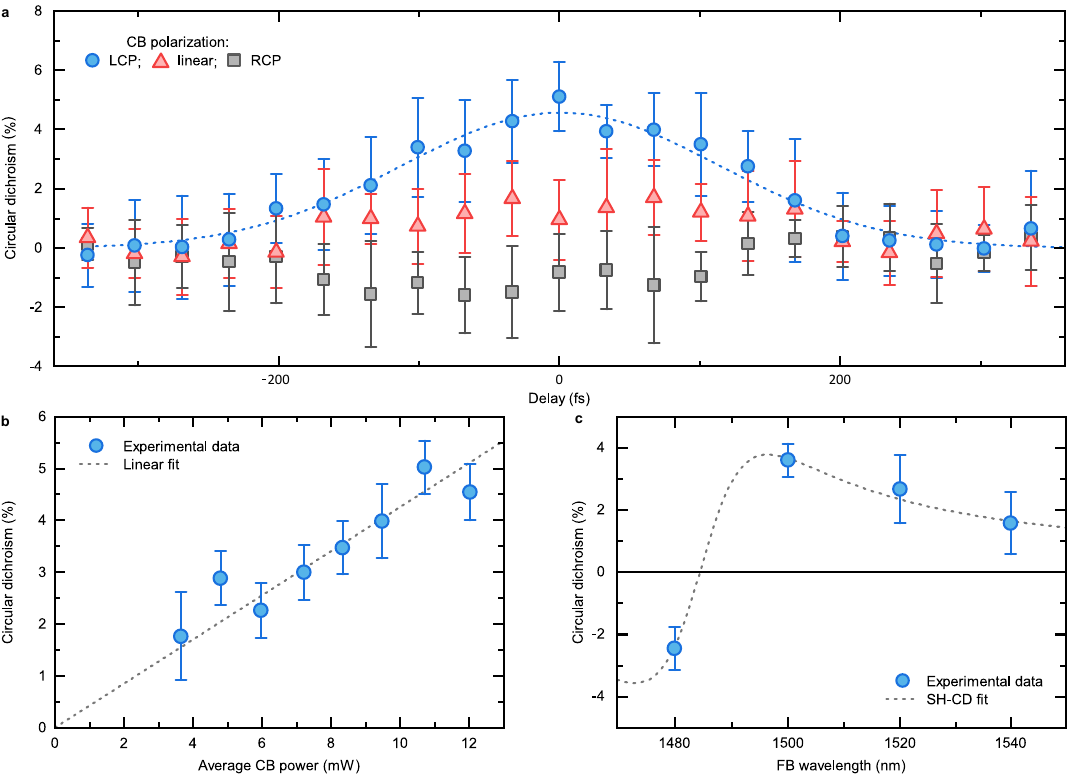}
    \caption{\textbf{Experimental tuning of the SH-CD.} \textbf{a}, Delay scans of the SH-CD for different CB polarizations (circular and linear). CB and FB wavelengths are \SI{1800}{nm} and \SI{1500}{nm}, respectively. The SH-CD is independent of time-delay when the CB is linearly polarized (red triangles), while it changes with time-delay and it flips sign when the helicity of the circularly polarized CB is reversed. The dashed line represents a Gaussian fit with full-width at half maximum of $\qty{286(21)}{\fs}$. The offset for large delay is set to be zero. \textbf{b}, CB power dependence of the SH-CD at zero time delay and for CB having LCP. The FB wavelength is fixed at \SI{1500}{\nm}. Intercept of the linear fit (dashed grey line) is set to zero. \textbf{c}, FB wavelength dependence of the SH-CD. Equation~\eqref{e2b} is used to fit the data using the experimental parameters  $\ezero = \SI{0.2}{\V \per \nm}$, $\Delta = \SI{1.662}{\eV}+\SI{8.2}{\meV}$ shift from optical dressing of the FB, and $\nu = 0.45$, leading to $T_2 = \qty{50(13)}{\fs}$ and $\abs{\Omega(\pm\text{K})}=\qty{8(2)}{\angstrom\squared}$.}
    \label{Fig4}
\end{figure}

With this experimental scheme, we can further modulate the SH-CD by tuning the CB to FB time delay and the CB intensity (Fig.~\ref{Fig4}). Based on equation~\eqref{e2b}, this provides a possible route towards the realization of ultrafast all-optical devices combining the valley degree of freedom with the engineering and read-out of the local Berry curvature. Panel a of Fig.~\ref{Fig4} shows the ultrafast dynamical change of the SH-CD, which occurs on a time-scale defined only by the CB and FB pulse duration ($\leq$ \SI{300}{\fs}). This ultrafast response time is inherent to the approach used for TRS breaking with below-gap off-resonant excitation. The sign change of the SH-CD by reversing the CB helicity corresponds to a change in the energy offset between the $\pm$K valleys, as predicted also by the linear dependence with $\sigma$ in equation~\eqref{e2b}. Furthermore, the SH-CD is independent of time delay for linear polarization of the CB (red triangles in Fig.~\ref{Fig4}~a), namely when TRS is preserved. 
A further experimental tuning knob is the CB intensity: coherent off-resonant bandgap modulation produces an asymmetry in the energy bandgap of the $\pm$K valleys that is linearly proportional to the light intensity, leading to the $\ezero^2$ dependence in equation~\eqref{e2b} and in agreement with the experimental results in Fig.~\ref{Fig4}~b.
Finally, the experimental SH-CD as a function of the FB wavelength exhibits a sign reversal close to half the equilibrium band gap (Fig.~\ref{Fig4}~c), although the circular polarization of the CB is kept fixed. This behavior is due to tuning of the SH wavelength across the light-dressed $\pm$K valleys, resulting in a modulation of the corresponding resonant contributions. Notably, this represents the TRS-broken counterpart of the effect reported in Ref.~\cite{Klimmer.2026} for linearly polarized excitation. Within our analytical framework, this mechanism is captured by the frequency-dependent function $f(\omega)$ in equation~\eqref{e2b}. The best fit of equation~\eqref{e2b} to our experimental data (grey dashed line in Fig.~\ref{Fig4}~c) provides a direct estimate of the dephasing time $T_2 = \qty{50(13)}{\fs}$ and of the $\pm$K resonant Berry curvature $\qty{8(2)}{\angstrom\squared}$, in good agreement with reported values from tight-binding~\cite{Cho.2018} and density functional theory~\cite{Xiao.2015} simulations.

\subsection{Summary} 
We have shown that nonlinear circular dichroism is a powerful ultrafast and all-optical method to  reveal the local Berry curvature at optical resonances. Conservation of angular momentum always imposes that the momentum transfer from the electromagnetic field to the crystal lattice is quantized in multiple integers of $r\hbar$, where $r$ is the discrete rotational frequency of the crystal. In addition to this fundamental conservation law, we demonstrate that the optically probed Berry curvature underlies the efficiency at which photons with opposite helicity are generated, resulting in a net nonlinear circular dichroism. 
We derived an analytical expression of this general model for a prototype atomically thin semiconductor with 3-fold rotational symmetry, focusing in particular on its SH-CD upon breaking of TRS with light-dressed states. This provides physical insights into the microscopic origin of the elements of the nonlinear optical susceptibility, and their link to the local quantum geometry. 
Finally, our analytical model  qualitatively and quantitatively describes experimental results of the dynamical second-harmonic circular dichroism in a monolayer WSe$_2$ sample, performed in a pump-probe configuration where the time delay, CB intensity and helicity, and FB wavelength are used as tuning knobs. 
This work offers a novel approach for the nonlinear all-optical read-out of the local quantum geometry, and it paves the way to a new era of ultrafast all-optical devices combining the spin and valley degrees of freedom to the local Berry curvature, thus advancing the field of ultrafast all-optical valleytronics~\cite{Gindl.2025, Gucci.2026, Tyulnev.2024, Mitra.2024, Seyler.2026}.


\section{Main text references}
\bibliographystyle{naturemag}


\begin{thebibliography}{10}
\expandafter\ifx\csname url\endcsname\relax
  \def\url#1{\texttt{#1}}\fi
\expandafter\ifx\csname urlprefix\endcsname\relax\def\urlprefix{URL }\fi
\providecommand{\bibinfo}[2]{#2}
\providecommand{\eprint}[2][]{\url{#2}}

\bibitem{Bloembergen.1980}
\bibinfo{author}{Bloembergen, N.}
\newblock \bibinfo{title}{Conservation laws in nonlinear optics*}.
\newblock \emph{\bibinfo{journal}{JOSA}} \textbf{\bibinfo{volume}{70}}, \bibinfo{pages}{1429} (\bibinfo{year}{1980}).

\bibitem{Simon.1968}
\bibinfo{author}{Simon, H.~J.} \& \bibinfo{author}{Bloembergen, N.}
\newblock \bibinfo{title}{Second-harmonic light generation in crystals with natural optical activity}.
\newblock \emph{\bibinfo{journal}{Phys. Rev.}} \textbf{\bibinfo{volume}{171}}, \bibinfo{pages}{1104--1114} (\bibinfo{year}{1968}).

\bibitem{Toftul.2023}
\bibinfo{author}{Toftul, I.} \emph{et~al.}
\newblock \bibinfo{title}{Nonlinearity-induced optical torque}.
\newblock \emph{\bibinfo{journal}{Phys. Rev. Lett.}} \textbf{\bibinfo{volume}{130}}, \bibinfo{pages}{243802} (\bibinfo{year}{2023}).

\bibitem{Li.2016}
\bibinfo{author}{Li, G.}, \bibinfo{author}{Zentgraf, T.} \& \bibinfo{author}{Zhang, S.}
\newblock \bibinfo{title}{Rotational doppler effect in nonlinear optics}.
\newblock \emph{\bibinfo{journal}{Nat. Phys.}} \textbf{\bibinfo{volume}{12}}, \bibinfo{pages}{736--740} (\bibinfo{year}{2016}).

\bibitem{Sodemann.2015}
\bibinfo{author}{Sodemann, I.} \& \bibinfo{author}{Fu, L.}
\newblock \bibinfo{title}{Quantum nonlinear hall effect induced by berry curvature dipole in time-reversal invariant materials}.
\newblock \emph{\bibinfo{journal}{Phys. Rev. Lett.}} \textbf{\bibinfo{volume}{115}}, \bibinfo{pages}{216806} (\bibinfo{year}{2015}).

\bibitem{Ma.2019}
\bibinfo{author}{Ma, Q.} \emph{et~al.}
\newblock \bibinfo{title}{Observation of the nonlinear hall effect under time-reversal-symmetric conditions}.
\newblock \emph{\bibinfo{journal}{Nature}} \textbf{\bibinfo{volume}{565}}, \bibinfo{pages}{337--342} (\bibinfo{year}{2019}).

\bibitem{Kang.2019}
\bibinfo{author}{Kang, K.}, \bibinfo{author}{Li, T.}, \bibinfo{author}{Sohn, E.}, \bibinfo{author}{Shan, J.} \& \bibinfo{author}{Mak, K.~F.}
\newblock \bibinfo{title}{Nonlinear anomalous hall effect in few-layer {WTe}$_2$}.
\newblock \emph{\bibinfo{journal}{Nat. Mater.}} \textbf{\bibinfo{volume}{18}}, \bibinfo{pages}{324--328} (\bibinfo{year}{2019}).

\bibitem{Soavi2025}
\bibinfo{author}{Soavi, G.} \& \bibinfo{author}{Wilhelm, J.}
\newblock \bibinfo{title}{{The role of Berry curvature derivatives in the optical activity of time-invariant crystals}} (\bibinfo{year}{2025}).
\newblock \eprint{arXiv:2501.03684}.

\bibitem{Xiao2010}
\bibinfo{author}{Xiao, D.}, \bibinfo{author}{Chang, M.-C.} \& \bibinfo{author}{Niu, Q.}
\newblock \bibinfo{title}{Berry phase effects on electronic properties}.
\newblock \emph{\bibinfo{journal}{Rev. Mod. Phys.}} \textbf{\bibinfo{volume}{82}}, \bibinfo{pages}{1959--2007} (\bibinfo{year}{2010}).

\bibitem{Schuler.2020}
\bibinfo{author}{Sch{\"u}ler, M.} \emph{et~al.}
\newblock \bibinfo{title}{Local berry curvature signatures in dichroic angle-resolved photoelectron spectroscopy from two-dimensional materials}.
\newblock \emph{\bibinfo{journal}{Sci. Adv.}} \textbf{\bibinfo{volume}{6}}, \bibinfo{pages}{eaay2730} (\bibinfo{year}{2020}).

\bibitem{Beaulieu.2024}
\bibinfo{author}{Beaulieu, S.} \emph{et~al.}
\newblock \bibinfo{title}{Berry curvature signatures in chiroptical excitonic transitions}.
\newblock \emph{\bibinfo{journal}{Sci. Adv.}} \textbf{\bibinfo{volume}{10}}, \bibinfo{pages}{eadk3897} (\bibinfo{year}{2024}).

\bibitem{Kim.2025}
\bibinfo{author}{Kim, S.} \emph{et~al.}
\newblock \bibinfo{title}{Direct measurement of the quantum metric tensor in solids}.
\newblock \emph{\bibinfo{journal}{Science (New York, N.Y.)}} \textbf{\bibinfo{volume}{388}}, \bibinfo{pages}{1050--1054} (\bibinfo{year}{2025}).

\bibitem{Kang.2025}
\bibinfo{author}{Kang, M.} \emph{et~al.}
\newblock \bibinfo{title}{Measurements of the quantum geometric tensor in solids}.
\newblock \emph{\bibinfo{journal}{Nat. Phys.}} \textbf{\bibinfo{volume}{21}}, \bibinfo{pages}{110--117} (\bibinfo{year}{2025}).

\bibitem{Sidilkover.2025}
\bibinfo{author}{Sidilkover, I.} \emph{et~al.}
\newblock \bibinfo{title}{Reexamining circular dichroism in photoemission from a topological insulator}.
\newblock \emph{\bibinfo{journal}{Phys. Rev. Res.}} \textbf{\bibinfo{volume}{7}} (\bibinfo{year}{2025}).

\bibitem{Herrmann2024}
\bibinfo{author}{Herrmann, P.} \emph{et~al.}
\newblock \bibinfo{title}{Nonlinear valley selection rules and all-optical probe of broken time-reversal symmetry in monolayer {WSe}$_2$}.
\newblock \emph{\bibinfo{journal}{Nat. Photon.}} \textbf{\bibinfo{volume}{19}}, \bibinfo{pages}{300--306} (\bibinfo{year}{2025}).

\bibitem{Sie.2017}
\bibinfo{author}{Sie, E.~J.} \emph{et~al.}
\newblock \bibinfo{title}{Large, valley-exclusive bloch-siegert shift in monolayer {WS}$_2$}  (\bibinfo{year}{2017}).

\bibitem{Cho.2018}
\bibinfo{author}{Cho, S.} \emph{et~al.}
\newblock \bibinfo{title}{Experimental observation of hidden berry curvature in inversion-symmetric bulk {2H-WSe}$_2$}.
\newblock \emph{\bibinfo{journal}{Phys. Rev. Lett.}} \textbf{\bibinfo{volume}{121}}, \bibinfo{pages}{186401} (\bibinfo{year}{2018}).

\bibitem{Xiao.2015}
\bibinfo{author}{Xiao, J.} \emph{et~al.}
\newblock \bibinfo{title}{Nonlinear optical selection rule based on valley-exciton locking in monolayer {WS}$_2$}.
\newblock \emph{\bibinfo{journal}{Light-Sci. Appl.}} \textbf{\bibinfo{volume}{4}}, \bibinfo{pages}{e366--e366} (\bibinfo{year}{2015}).

\bibitem{Wang.2018}
\bibinfo{author}{Wang, G.} \emph{et~al.}
\newblock \bibinfo{title}{Colloquium: Excitons in atomically thin transition metal dichalcogenides}.
\newblock \emph{\bibinfo{journal}{Rev. Mod. Phys.}} \textbf{\bibinfo{volume}{90}}, \bibinfo{pages}{021001} (\bibinfo{year}{2018}).

\bibitem{Man.2021}
\bibinfo{author}{Man, M. K.~L.} \emph{et~al.}
\newblock \bibinfo{title}{Experimental measurement of the intrinsic excitonic wave function}.
\newblock \emph{\bibinfo{journal}{Sci. Adv.}} \textbf{\bibinfo{volume}{7}}, \bibinfo{pages}{eabg0192} (\bibinfo{year}{2021}).

\bibitem{Yao2008}
\bibinfo{author}{Yao, W.}, \bibinfo{author}{Xiao, D.} \& \bibinfo{author}{Niu, Q.}
\newblock \bibinfo{title}{Valley-dependent optoelectronics from inversion symmetry breaking}.
\newblock \emph{\bibinfo{journal}{Phys. Rev. B}} \textbf{\bibinfo{volume}{77}}, \bibinfo{pages}{235406} (\bibinfo{year}{2008}).

\bibitem{Cao2012}
\bibinfo{author}{Cao, T.} \emph{et~al.}
\newblock \bibinfo{title}{Valley-selective circular dichroism of monolayer molybdenum disulphide}.
\newblock \emph{\bibinfo{journal}{Nat. Commun.}} \textbf{\bibinfo{volume}{3}}, \bibinfo{pages}{887} (\bibinfo{year}{2012}).

\bibitem{Taghizadeh2019}
\bibinfo{author}{Taghizadeh, A.} \& \bibinfo{author}{Pedersen, T.~G.}
\newblock \bibinfo{title}{Nonlinear optical selection rules of excitons in monolayer transition metal dichalcogenides}.
\newblock \emph{\bibinfo{journal}{Phys. Rev. B}} \textbf{\bibinfo{volume}{99}}, \bibinfo{pages}{235433} (\bibinfo{year}{2019}).

\bibitem{Friedrich.2026}
\bibinfo{author}{Friedrich, F.} \emph{et~al.}
\newblock \bibinfo{title}{Measurement of optically induced broken time-reversal symmetry in atomically thin crystals}.
\newblock \emph{\bibinfo{journal}{Nat. Photon.}} \textbf{\bibinfo{volume}{20}}, \bibinfo{pages}{186--193} (\bibinfo{year}{2026}).

\bibitem{Boyd.2020}
\bibinfo{author}{Boyd, R.~W.}
\newblock \emph{\bibinfo{title}{Nonlinear Optics / Robert W. Boyd}} (\bibinfo{publisher}{{Academic Press}}, \bibinfo{address}{London}, \bibinfo{year}{2020}), \bibinfo{edition}{4th ed.} edn.

\bibitem{RodriguezCarvajal.2012}
\bibinfo{author}{Rodr{\'i}guez-Carvajal, J.} \& \bibinfo{author}{Bour{\'e}e, F.}
\newblock \bibinfo{title}{Symmetry and magnetic structures}.
\newblock \emph{\bibinfo{journal}{EPJ Web of Conferences}} \textbf{\bibinfo{volume}{22}}, \bibinfo{pages}{00010} (\bibinfo{year}{2012}).

\bibitem{Herrmann.2023}
\bibinfo{author}{Herrmann, P.} \emph{et~al.}
\newblock \bibinfo{title}{Nonlinear all-optical coherent generation and read-out of valleys in atomically thin semiconductors}.
\newblock \emph{\bibinfo{journal}{Small}} \textbf{\bibinfo{volume}{19}}, \bibinfo{pages}{e2301126} (\bibinfo{year}{2023}).

\bibitem{Klimmer.2026}
\bibinfo{author}{Klimmer, S.} \emph{et~al.}
\newblock \bibinfo{title}{Probing ultrafast coherent bandgap modulation in monolayer {WSe}$_2$ by nonlinear optics}.
\newblock \emph{\bibinfo{journal}{Adv. Opt. Mater.}} \textbf{\bibinfo{volume}{14}} (\bibinfo{year}{2026}).

\bibitem{Gindl.2025}
\bibinfo{author}{Gindl, A.}, \bibinfo{author}{{\v{C}}mel, M.}, \bibinfo{author}{Troj{\'a}nek, F.}, \bibinfo{author}{Mal{\'y}, P.} \& \bibinfo{author}{Koz{\'a}k, M.}
\newblock \bibinfo{title}{Ultrafast room-temperature valley manipulation in silicon and diamond}.
\newblock \emph{\bibinfo{journal}{Nat. Phys.}} \textbf{\bibinfo{volume}{21}}, \bibinfo{pages}{947--952} (\bibinfo{year}{2025}).

\bibitem{Gucci.2026}
\bibinfo{author}{Gucci, F.} \emph{et~al.}
\newblock \bibinfo{title}{Encoding and manipulating ultrafast coherent valleytronic information with lightwaves}.
\newblock \emph{\bibinfo{journal}{Nat. Photon.}}  (\bibinfo{year}{2026}).

\bibitem{Tyulnev.2024}
\bibinfo{author}{Tyulnev, I.} \emph{et~al.}
\newblock \bibinfo{title}{Valleytronics in bulk {MoS}$_2$ with a topologic optical field}.
\newblock \emph{\bibinfo{journal}{Nature}} \textbf{\bibinfo{volume}{628}}, \bibinfo{pages}{746--751} (\bibinfo{year}{2024}).

\bibitem{Mitra.2024}
\bibinfo{author}{Mitra, S.} \emph{et~al.}
\newblock \bibinfo{title}{Light-wave-controlled haldane model in monolayer hexagonal boron nitride}.
\newblock \emph{\bibinfo{journal}{Nature}} \textbf{\bibinfo{volume}{628}}, \bibinfo{pages}{752--757} (\bibinfo{year}{2024}).

\bibitem{Seyler.2026}
\bibinfo{author}{Seyler, K.~L.} \emph{et~al.}
\newblock \bibinfo{title}{Valleytronics in 2d materials roadmap} (\bibinfo{year}{2026}).

\end{thebibliography}

\section{Methods}
\subsection{Second-harmonic circular dichroism measurements}
For the two-color SH-CD experiments, the CB and FB are taken from two synchronized optical parametric oscillators (Levante IR fs, A.P.E) pumped by an Ytterbium-doped laser with central wavelength at \SI{1030}{nm}, repetition rate \SI{76}{\mega\Hz}, and pulse duration  $\sim$\SI{100}{fs} (FLINT FL2-12, LIGHT CONVERSION). The time delay between CB and FB is tuned by a motorized delay stage (M-414.2PD, PI with controller C-863.12, PI). The FB polarization and power are controlled by a combination of a linear (WP25M-UB, Thorlabs) and a Glan-Thompson polarizer (GTH10M). The power of the CB is controlled by the combination of two linear polarizers (WP25M-UB, Thorlabs). The polarization of the CB is tuned by combining a half-wave plate (HWP) and quarter-wave plate (QWP, RSU 1.2.15 and 1.4.15, B-Halle) in two motorized rotational mounts (PRM1/MZ8, Thorlabs). The CB and FB are subsequently combined at a Polkadot beamsplitter (BPD5254-G01, Thorlabs), and focused onto the sample with a 40x reflective objective (LMM-40X-UVV, Thorlabs). 
After the sample, we use a lens (C330TMD, Thorlabs) and different sets of spectral filters to isolate the SH signal from the residual transmitted CB and FB. To measure the ellipticity and CD of the SH beam, we use a QWP (\#63-935, Edmund optics) at $\pm$\SI{45}{\degree} with respect to a fixed linear polarizer (WP25M-UB, Thorlabs), that is aligned parallel to the SH of the FB.  
The spectra in Fig.~\ref{Fig3} are taken with a monochromator (iHR320, HORIBA) combined with a nitrogen cooled silicon detector (Symphony II, HORIBA), and the FB is blocked by a shortpass filter (FESH0950, Thorlabs). For the measurements in Fig.~\ref{Fig4}, we detect the SH signal with a silicon avalanche photodiode (APD440A, Thorlabs) and lock-in amplification (MFLI, Zurich Instruments), which filters the signal coming from the FB modulated at \SI{967}{\Hz} via an optical chopper (MC2000B, Thorlabs). Bandpass filters are chosen for each specific SH wavelength (FBHxxx-10, Thorlabs). 

\subsection{Sample preparation and characterization}
The monolayer WSe$_2$ was mechanically exfoliated  from a bulk crystal (HQ graphene) onto polydimethylsiloxane and transferred onto a fused silica substrate using a commercial transfer stage (HQ graphene). The monolayer nature of the sample was determined by photoluminescence (PL) measurements and a cross-polarized SH polarization pattern with linearly polarized FB was performed to rule out the presence of strain in the sample (see Supporting Information, Sec.~S3).

\subsection{Experimental error estimation}
The error bars in Fig.~\ref{Fig4}a are based on error propagation of the standard deviation over 5 acquisitions for each data point. The SH-CD values in Fig.~\ref{Fig4}b and c are the amplitudes of the Gaussian fits of time delay measurements like in Fig.~\ref{Fig4}a. The error bars are the fit errors of the amplitude of the Gaussian fits. The Gaussian fit algorithm considers the standard deviation for each point of the delay curve. For the estimation of $\abs{\Omega(\pm\text{K})}$ and $T_2$ we use the experimental values including the error bars in Fig.~\ref{Fig4}b and fit equation~\eqref{e2b} to it. The deviation given for the values in the caption of Fig.~\ref{Fig4} are the standard deviations extracted from the covariance matrix of the fit parameters.

\section{Author contribution}
G.S. and J.W. conceived the work. N.T. and P.H. prepared the sample, performed the measurements, and analyzed the data. N.T., P.H., and G.S. interpreted the experimental results. C.S., F.E., and J.W. developed the analytical model. The manuscript was written with contributions from all co-authors.

\section{Acknowledgements}
We thank Adrian Seith and Shridhar Shanbhag for helpful discussions. The authors thank Sebastian Klimmer for valuable assistance with figure design and visualization 
G.S. acknowledges funding from the Deutsche Forschungsgemeinschaft (DFG, German Research Foundation) via the SFB 1375 (project number 398816777, subprojects B5 and C4), IRTG 2675 (project number 437527638) and WHAT-A-TWIST (project number 547611111).
J.W. acknowledges the DFG for funding via the Emmy Noether Programme (project number 503985532), CRC 1277 (project number 314695032, subproject A03) and RTG 2905 (project number 502572516).
F.E.~acknowledges that
this work was supported by the  DFG under EV30/12-1, EV30/14-1, EV30-16-1, further CRC 1277
(Project-ID 314695032, subproject A03), and RTG 2905
(project number 502572516).

\section{Additional information}
Correspondence and requests for materials should be addressed to Jan Wilhelm and Giancarlo Soavi.

\newpage


\end{document}



\section{Supplementary Information}
Nonlinear Circular Dichroism Reveals the Local Berry Curvature

\section{Author list}
Nele Tornow$^{1}$, Paul Herrmann$^{1}$, Clemens Schneider$^{2}$, Ferdinand Evers$^{2,3}$, Jan Wilhelm$^{2,\ddagger}$, and {Giancarlo} Soavi$^{1,4,\star}$

\section{Affiliations}
\noindent
$^1$Institute of Solid State Physics, Friedrich Schiller University Jena, Helmholtzweg 5, 07743 Jena, Germany
\newline
$^2$Institute of Theoretical Physics and Regensburg Center for Ultrafast Nanoscopy (RUN), University of Regensburg, Universitätsstrasse 31, 93053 Regensburg, Germany
\newline
$^3$Halle-Berlin-Regensburg Cluster of Excellence CCE, Universitätsstrasse 31,  93053 Regensburg, Germany
\newline
$^4$Abbe Center of Photonics, Friedrich Schiller University Jena, Albert-Einstein-Straße 6, 07745 Jena, Germany
\newline
$^{\ddagger}$ jan.wilhelm@physik.uni-regensburg.de
$^{\star}$ giancarlo.soavi@uni-jena.de

\maketitle

\newpage

\section{S1 Nonlinear circular dichroism and Berry curvature in the limit of a single resonant crystal momentum}
We consider the case of a single resonant crystal momentum with the $m$th harmonic frequency, $\varepsilon_{cv}(\kres)\eqt m\hbar\omega$ and $\varepsilon_{cv}(\bk)\neqt m\hbar\omega$ for all other $\bk\in \text{BZ}/\{\kres\}$.
Our analysis starts from the definition of CD in the $m$th harmonic
\begin{align}
    \text{$m$H-CD} = \frac{|j_+|^2-|j_-|^2}{|j_+|^2+|j_-|^2} \label{e15a}
\end{align}
using the circular current density oscillating at frequency~$m\omega$~\cite{Wilhelm2021},
\begin{align}
&   j_\pm \coloneqq \frac{1}{\sqrt{2}}\big[j_x(m\omega)\pm i j_y(m\omega)\big]\;,\\[0.5em]
 &  \bj(m\omega) \coloneqq \frac{q}{V} \,\text{Tr}[\bv \rho(m\omega)] = q \intbzdkpitwopi \sum_{nn'}\bv_{nn'}(\bk)\, \rho_{n'n}(\bk,m\omega) \,,\label{e16a}
\end{align}
where $q$ is the electron charge, $V$ the unit cell volume, $\bv$ the velocity operator, $\rho(t)$ the one-electron density matrix and $d$ the dimension.
The trace includes the Brillouin Zone (BZ) integration and summation of the band indices~$n,n'$. 
In the perturbative regime, the absolute value of the density matrix  $|\rho_{cv}(\bk,m\omega)|$ has a peak at $\kres$, so we evaluate the BZ integral in equation~\eqref{e16a} as $\bk$-width of the peak,~$W$, times the height (maximum of the integrand at $\kres$), 
\begin{align}
  \bj(m\omega) = \frac{q\,W}{(2\pi)^d} \,\bv_{vc}(\kres)\, \rho_{cv}(\kres,m\omega) \,, \label{e17}
\end{align}

Inserting equation~\eqref{e17} into equation~\eqref{e15a} leads to a circular dichroism
\begin{align}
    \text{$m$H-CD} = \frac{|v_+|^2-|v_-|^2}{|v_+|^2+|v_-|^2}\label{e16}
\end{align}
with the definition
\begin{align}
    v_\pm \coloneqq \frac{1}{\sqrt{2}} \big(v_x \pm i v_y\big)\;,\;\;v_\gamma \coloneqq v_{vc}^\gamma(\kres) \,.
\end{align}
The expression in equation~\eqref{e15a}/\eqref{e16} shows that the $m^{th}$ circular dichroism reflects the sense of rotation of the $m^{th}$ harmonic interband current~$\bj$ caused by the interband motion at the resonant $\bk$-point. 
The sign of $|v_+|^2\mt|v_-|^2\eqt 2\,\text{Im}[v_xv_y^*]$ determines whether the interband current rotates left- or right-handed, corresponding to a preferential emission of $\sigma^+$ or $\sigma^-$ light. 
Using $\Omega_z\eqt2\,\varepsilon_{cv}^{-2}\,\mathrm{Im}[v_xv_y^*]$, the dichroism thus measures the Berry curvature at $\kres$ normalized by the total transition strength~$d_{vc}^\gamma(\bk)\eqt iv_{vc}^\gamma(\bk)/\varepsilon_{cv}(\bk)$:
\begin{align}
    \text{$m$H-CD} &
     = \frac{ \Omega(\kres) }{ ||\mathbf{d}_{vc}(\kres)||^2} \,.
     \label{e21}
\end{align}
Indeed, nonzero $\mathrm{Im}(v_xv_y^*)$ implies that the $x$ and $y$ components of the optical current are phase-shifted, causing the excited electron–hole polarization to follow a circular (or elliptical) trajectory in time; the sign of this rotation is set by $\mathrm{sgn}[\Omega_z(\kres)]$. 
In this way, $m$H-CD at a single resonant $\bk$ probes the  angular momentum of the interband velocity and thereby the local Berry curvature of the band structure.
\newpage
\section{S2 Microscopic model of circular dichroism in second harmonic generation for a TMD monolayer} \label{SI_sec4}

For a theoretical analysis of the SH-CD, we start from 
\begin{align}
    \text{SH-CD} =
   \frac{ \left|\chi_{+\alpha\alpha}\right|^2-
    \left|\chi_{-\alpha\alpha}\right|^2}{\left|\chi_{+\alpha\alpha}\right|^2+
    \left|\chi_{-\alpha\alpha}\right|^2} \label{e1a}
\end{align}
where $\chi_{\sigma\alpha\alpha} $ is the second-order nonlinear optical susceptibilities for incoming linear polarization along the $\alpha$-direction and second-harmonic output with circular polarization $\sigma\eqt{\pm} 1$:
\begin{align}
\chi_{\sigma\alpha\alpha}
=
\frac{1}{\sqrt{2}}\left(
\chi_{x\alpha\alpha} + i\sigma  \chi_{y\alpha\alpha} \right)\,.\label{e03}
\end{align}
Using this definition, the SH-CD~\eqref{e1a} simplifies to
\begin{align}
    \text{SH-CD} =
   \frac{  2\,  \text{Im}  \left(\chi_{x\alpha\alpha}
\,\chi_{y\alpha\alpha}^*  \right)  }{|\chi_{x\alpha\alpha} |^2+
|\chi_{y\alpha\alpha} |^2  } \label{e3a}\,.
\end{align}
Note that equations~\eqref{e1a} and~\eqref{e3a} are valid only for materials with broken space-inversion symmetry, since in centrosymmetric media the second-order susceptibility $\chi_{\alpha\beta\gamma}$ vanishes.

An expression for $\chi_{\gamma\alpha\alpha}$ ($\gamma\eqt x,y$) can be derived from perturbative solutions of the semiconductor Bloch equations~\cite{Aversa1995,Herrmann2024} using a two-band model with valence band~$v$ and conduction band~$c$. We start from equation~(25) in the supporting information of Ref.~\cite{Herrmann2024},
\begin{align}
\chi_{\gamma\alpha\alpha}&=C \, \intbzdkpi\;  
\frac{v_{vc}^\gamma}
 {\varepsilon_{cv}-i\hbar/T_2-2\hbar\omega} \Bigg[\frac{\partial}{\partial k_\alpha}\,\frac{d_{cv}^\alpha }{\varepsilon_{cv}-i\hbar/T_2-\hbar\omega} 
 +i              
 \frac{d_{cv}^\alpha(d_{cc}^\alpha-d_{vv}^\alpha) }{\varepsilon_{cv}-i\hbar/T_2-\hbar\omega}
 \Bigg]
 \,,
 \label{e01}
\end{align}
where $C$ is a constant~\cite{Aversa1995}, $v_{vc}^\gamma(\bk)\eqt {-}id_{vc}^\gamma(\bk)\varepsilon_{cv}(\bk)$ is the velocity matrix element, \linebreak $\varepsilon_{cv}(\bk)\eqt\varepsilon_{c}(\bk)\mt\varepsilon_{v}(\bk) $ is the energy gap between valence and conduction bands at~$\bk$, and $T_2$ is the dephasing time. 
We suppress the $k$-dependencies of all quantities in equation~\eqref{e01}.
We first assume that TRS is preserved at equilibrium, such that driving addresses at least two $\pm\bk$-valleys (in case of TMDs: $\pm$\,K, labelled as~$\tau\eqt{\pm}1$).
Second, we focus on resonant driving at a band extremum, \textit{i.e.},  $\partial\varepsilon_{cv}/\partial\bk\eqt 0$, and the integral in equation~\eqref{e01} simplifies to a sum over the two resonant $k$-points:
\begin{align}
 \chi_{\gamma\alpha\alpha}  &= \tilde C  \sum_{\tau=\pm 1}
 W(\tau)\,f_\alpha(\tau)  \,d_{vc}^\gamma(\tau) 
\,,
 \label{e02}
 \\[0.5em]
 W(\tau) &= \frac{1}
 {\varepsilon_{cv}(\tau)-i\hbar/T_2-2\hbar\omega}\,,
\hspace{2em} 
 f_\alpha(\tau) = \left.\frac{\partial d_{cv}^\alpha }{\partial k_\alpha} \right|_\tau
 +i              
  {d_{cv}^\alpha(\tau)\big[d_{cc}^\alpha(\tau)-d_{vv}^\alpha(\tau)\big] }
\,.\label{e5}
\end{align}
Here, $\omega$ is the frequency of the fundamental beam which is tuned to the second-harmonic resonance of the $\pm$K points, $\varepsilon_{cv}(\tau)- 2\hbar\omega\approx 0$. The energy gap in the definition of $v_{vc}^\gamma$ is absorbed into the prefactor $\tilde C$, since the $\tau$-dependent part of $\varepsilon_{cv}(\tau)$ is already quadratic in the electric field (see equation~\eqref{e7a}).
To calculate the dipole matrix elements we use the TMD Hamiltonian~\cite{Friedrich.2026} for $\bk$ close to $\pm$K ($\tau\eqt \pm1$):
\begin{align}
\mathbf{h}(\bk)=
\begin{pmatrix}
\frac{1}{2}\varepsilon_{cv}(\tau) +\mathcal{O}(\kappa^2)& \gamma^* f^*(\bk) \\
\gamma f(\bk) & -\frac{1}{2}\varepsilon_{cv}(\tau) + \mathcal{O}(\kappa^2)
\end{pmatrix} \label{c1}
\end{align}
with
\begin{align}
f(\bk) = -(i\kappa_x + \kappa_y\tau) +\zeta(\kappa_x + i\kappa_y\tau)^2 + \mathcal{O}(\kappa^3)\label{c5}
\end{align}
where $\gamma$ is the effective hopping, $\kappa_{x(y)}=a(k_{x(y)}-\text{K}_{x(y)})$ is the dimensionless wave vector relative to the $\pm$K points, $a$ is the lattice constant and $\zeta=\frac{\sqrt{3}}{12}$. Note that the $\pm$K points are parametrised by $\tau$ in the following way: $\pm \text{K}=\frac{2\pi}{3a}\left( \sqrt{3}, \tau \right)$.
The gap~$\varepsilon_{cv}(\tau)$ at the $\pm$K valley differs because of the optical Stark and Bloch-Siegert shift~\cite[equation~(5)]{Sie.2017}\cite{Herrmann2024},
\begin{align}
\varepsilon_{cv}(\tau) = \Delta + \frac{d_0^2\,\ezero^2}{\Delta  +\sigma\tau\omega_\text{CB} }= \Delta + \frac{d_0^2\,\ezero^2}{\Delta}\,\frac{1}{1  +\nu\sigma\tau }  \,,\label{e7a}
\end{align}
where $\Delta$ is the equilibrium optical gap at the $\pm$K points, and $\ezero$, $\sigma$, $\omega_\text{CB}$ are  the field strength, helicity and frequency of the CB. 
We also introduce the prefactor~$\nu$ via $\omega_\text{CB}\eqt \nu\Delta$.
Note that TRS is broken in equation~\eqref{e7a} as a consequence of the control laser field. 

After diagonalising $\mathbf{h}$~\eqref{c1}, we use the eigenstates $\ket{v\bk}$ and $\ket{c\bk}$ to get the relevant dipole matrix elements~\cite[supplementary information, equation~(49)]{Friedrich.2026}:
\begin{align}
d_{vc}^x&=iq\bra{v\bk}\frac{\partial}{\partial k_x} \ket{c\bk}=\frac{aq\gamma}{\varepsilon_{cv}(\tau)} + \mathcal{O}(\kappa) \label{c2} \\
d_{vc}^y&=iq\bra{v\bk}\frac{\partial}{\partial k_y} \ket{c\bk}=-i\tau\frac{aq\gamma}{\varepsilon_{cv}(\tau)} + \mathcal{O}(\kappa) \\
d_{cv}^x&=iq\bra{c\bk}\frac{\partial}{\partial k_x} \ket{v\bk}=\frac{aq\gamma^*}{\varepsilon_{cv}(\tau)}(1 -2i\zeta\kappa_x - 2\zeta\tau\kappa_y) + \mathcal{O}(\kappa^2) \\
d_{cv}^y&=iq\bra{c\bk}\frac{\partial}{\partial k_y} \ket{v\bk}=\frac{aq\gamma^*}{\varepsilon_{cv}(\tau)}(i\tau -2\zeta\tau\kappa_x + 2i\zeta\tau\kappa_y) + \mathcal{O}(\kappa^2) \label{c3}
\end{align}
$q$ is the charge of an electron and $\zeta=\sqrt{3}/12$. Since $d_{vv}^\alpha$ and $d_{cc}^\alpha$ are of order $\kappa$, they can be neglected in further calculations. $d_0= |aq\gamma/\Delta|$ from equation~\eqref{e7a} is the dipole moment of the unperturbed material at the $\pm$K points.

%
We define the Berry curvature~\cite{Xiao2010}
\begin{align}
    \Omega (\tau)=
    2\,\text{Im}[ d_{vc}^x(\tau)\,d_{cv}^y(\tau)].\label{e10}
\end{align}
We then evaluate the numerator and denominator of CD-SHG~\eqref{e3a}:
\begin{align}
  \text{Im}  \left(\chi_{x\alpha\alpha}
\,\chi_{y\alpha\alpha}^*  \right) 
&\overset{\eqref{e02}}{=}
\,|\tilde C|^2\,\text{Im} \left[ 
\sum_{\tau=\pm 1}|W(\tau)|^2\,|f_\alpha(\tau)|^2\,d^x_{vc}(\tau)\,d^y_{cv}(\tau) 
\right]\nonumber
\\ &+
|\tilde C|^2\,\text{Im} \Big[ 
W(+)\,W^*(-) \, f_\alpha(+)\,f^*_\alpha(-)\,d^x_{vc}(+)\,d_{cv}^y(-)\nonumber
\\&
\hspace{3.7em}+
W^*(+)\,W(-) \, f_\alpha^*(+)\,f_\alpha(-)\hspace{-1.2em}
\underbrace{d^x_{vc}(-)\,d_{cv}^y(+)}_{\displaystyle 
 =\left[d^x_{vc}(+)\,d_{cv}^y(-)\right]^*}\hspace{-1em}
\Big]\nonumber
\\ &\hspace{-1.6em}\underset{\eqref{e10}}{\overset{\text{Im}(a+a^*)=0}{=}}\hspace{0.5em}\frac{|\tilde C|^2}{2}  
\sum_{\tau=\pm 1}|W(\tau)|^2\,|f_\alpha(\tau)|^2\,\Omega (\tau) 
 \,,
 \\[1em]
 |\chi_{x\alpha\alpha} |^2+
|\chi_{y\alpha\alpha} |^2  &
\overset{\eqref{e02}}{=}
\,|\tilde C|^2\,
\sum_{\tau=\pm 1}|W(\tau)|^2\,|f_\alpha(\tau)|^2\,||\mathbf{d}_{vc}(\tau)||^2 
\,.
\end{align}

\begin{figure}[h!]
\centering
\includegraphics{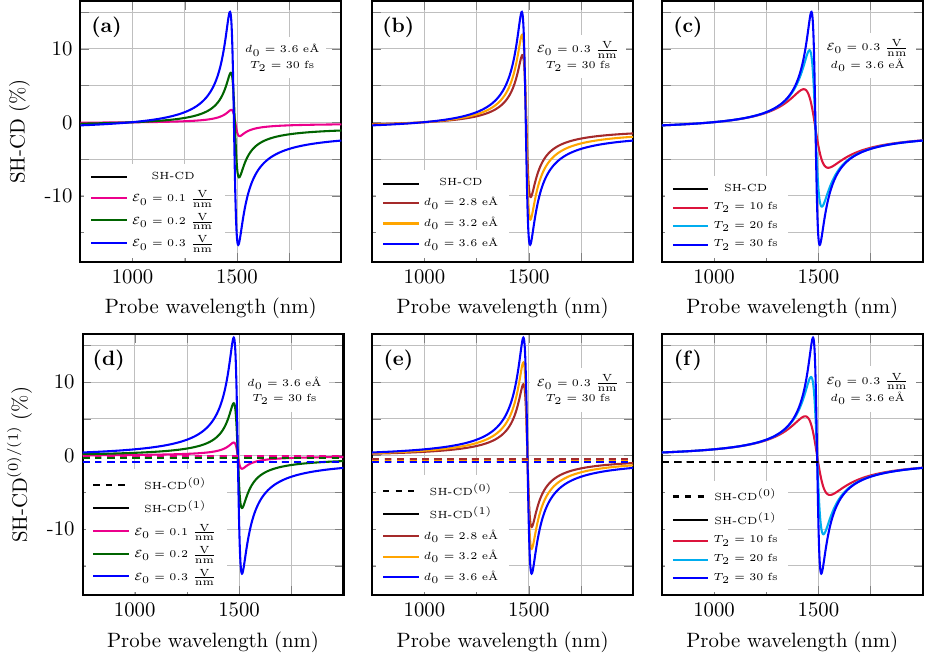}
\caption{\textbf{Analytical results for the SH-CD in monolayer WSe$_2$.} Dependence of SH-CD of a TMD monolayer on \textbf{(a)}/\textbf{(d)} the pump field strength~$\ezero$, \textbf{(b)}/\textbf{(e)} the dipole $d_0$, and \textbf{(c)}/\textbf{(f)} the dephasing time~$T_2$. \textbf{(a)} - \textbf{(c)} are the exact results from equation~\eqref{c4}, \textbf{(d)} - \textbf{(f)} show the approximated results from equation~\eqref{c11}. Throughout, we use $\Delta\eqt$1.66\,eV, $\nu\eqt0.415$ and $\sigma\eqt -1$.}
\label{SI_Fig1}
\end{figure}

We used
\begin{align}
d^x_{vc}(-)\,d_{cv}^y(+) =\frac{aq\gamma}{\varepsilon_{cv}(-)}\frac{iaq\gamma^*}{\varepsilon_{cv}(+)} = \left[\frac{aq\gamma}{\varepsilon_{cv}(+)}\frac{-iaq\gamma^*}{\varepsilon_{cv}(-)}\right]^*=\left[d^x_{vc}(+)\,d_{cv}^y(-)\right]^*,
\end{align}
and equation~\eqref{e3a} becomes
\begin{align}
    \text{SH-CD} =
   \frac{\sum\limits_{\tau=\pm 1}|W(\tau)|^2\,|f_\alpha(\tau)|^2\,\Omega (\tau) }{\sum\limits_{\tau=\pm 1}|W(\tau)|^2\,|f_\alpha(\tau)|^2\,||\mathbf{d}_{vc}(\tau)||^2 }. \label{c4}
\end{align}
We show the dependence of SH-CD on the fundamental beam frequency~$\omega$, field strength $\ezero$, dipole moment $d_0$ and dephasing $T_2$ in Fig.~\ref{SI_Fig1} (a)\,--\,(c), respectively (using band structure and dipole matrix elements from equations~\eqref{e7a} -~\eqref{c3}). 
Note that equation~\eqref{c4} simplifies to equation~\eqref{e21} in the extreme case of  $|W(+)|^2\,{\gg}\, |W(-)|^2$, for example using an extremely large control beam field amplitude~$\ezero$.

To further simplify equation~\eqref{c4}, we introduce a dimensionless variable 
\begin{align}
\lambda= \frac{d_0^2\,\ezero^2}{\Delta^2}\ll 1
\end{align}
which is small for the CB field strength~$\ezero$ used in the experiments. 
We then expand SH-CD in leading order of $\lambda$. From equation~\eqref{e7a}, we have
\begin{align}
\frac{1}{\varepsilon_{cv}^2(\tau)} = \frac{1}{\Delta^2}\left(1 - \frac{2\lambda}{1+ \nu\sigma\tau} + \ols \right)
&=\frac{1}{\Delta^2}\big[1- \lambda G+\tau\lambda F+\ols\big] \label{c10}
\end{align}
with $F\eqt {2\sigma\nu}/({1- \nu^2})$ and $G\eqt 2/(1-\nu^2)$.
Equation~\eqref{c10} and the dipole matrix elements from equations~\eqref{c2} -~\eqref{c3} can be employed to calculate $\Omega (\tau)$, $||\mathbf{d}_{vc}(\tau)||^2$ and $|f_\alpha(\tau)|^2$:
\begin{align}
\Omega(\tau)&=2\,\text{Im}[ d_{vc}^x(\tau)\,d_{cv}^y(\tau)]
=\frac{2\tau|\gamma|^2a^2q^2}{\varepsilon_{cv}^2(\tau)}
= \left[\tau-\tau\lambda G+\lambda F + \ols\right]\abs{\Omega(\pm \text{K})}
\label{c6} \\ 
||\mathbf{d}_{vc}(\tau)||^2&=\frac{2|\gamma|^2a^2q^2}{\varepsilon_{cv}^2(\tau)}= 2d_0^2\big[1- \lambda G+\tau\lambda F+\ols\big]\label{c7}\\
|f_\alpha(\tau)|^2&= \left|\left.\frac{\partial d_{cv}^\alpha }{\partial k_\alpha} \right|_\tau\right|^2= \frac{4|\gamma|^2a^4q^2\zeta^2}{\varepsilon_{cv}^2(\tau)}= A\big[1- \lambda G+\tau\lambda F+\ols\big] \label{c8}
\end{align}
with the constants $\abs{\Omega(\pm \text{K})}=2d_0^2={2|\gamma|^2a^2q^2}/{\Delta^2}$ and 
$A=4d_0^2a^2\zeta^2$.
Similarly, we expand $|W(\tau)|^2$:
\begin{align}
|W(\tau)|^2&=\frac{1}{|\varepsilon_{cv}(\tau){-}\frac{i\hbar}{T_2}{-}2\hbar\omega|^2}=
\frac{\left|{\eta}/{\Delta}\right|^2}{\left|1{+} \frac{\lambda\eta}{1{+}\nu\sigma\tau} \right|^2}
=
\left|\frac{\eta}{\Delta}\right|^2\big[1{-}\lambda G\text{Re}\,\eta{+}\tau\lambda F \text{Re}\,\eta{+} \ols\big] \label{c9}
\end{align}
with $\eta=1/({1-\frac{i\hbar}{T_2\Delta} -\frac{2\hbar\omega}{\Delta}})$.
Inserting equations~\eqref{c6}\,--\,\eqref{c9} into equation~\eqref{c4} yields
\begin{align}
  \text{SH-CD}&=\frac{ \abs{\Omega(\pm \text{K})}}{2d_0^2}\frac{\sum\limits_{\tau=\pm 1}\tau(1-\lambda G\text{Re}\,\eta+\tau\lambda F\, \text{Re}\,\eta)(1- \lambda G+\tau\lambda F)^2 +\ols}{\sum\limits_{\tau=\pm 1}(1-\lambda G\text{Re}\,\eta+\tau\lambda F\, \text{Re}\,\eta)(1- \lambda G+\tau\lambda F)^2+\ols}\\[0.5em]
  &=\frac{ \abs{\Omega(\pm \text{K})}}{2d_0^2}\lambda F( 2+ \text{Re}\,\eta(\omega))+\ols
 \eqqcolon \cdshgzero+\cdshgone(\omega)\label{c11}.
\end{align}
We observe that the SH-CD consists of a frequency-independent term $\cdshgzero$---which arises because of the difference of dipole matrix elements and Berry curvature at the $\pm$K valley--- and a  frequency-dependent term $\cdshgone$ which arises due to the different resonance condition at the $\pm$K valley via $\text{Re}\,\eta(\omega)$.
 Both terms are plotted in Fig.~\ref{SI_Fig1} (d)\,--\,(f) for different values of $\ezero$, $d_0$ and $T_2$, respectively. When the fundamental beam is not resonant, the approximated SH-CD matches the exact results from equation~\ref{c4}. However, the resonance in the exact result is shifted to higher frequencies compared to the approximative result (for $\sigma=1$). To see this effect in Fig.~\ref{SI_Fig1} (d)\,--\,(f), one needs to include higher orders of $\lambda$.\\ In the main text, we use the following notation for the SH-CD (equation~\eqref{c4}):
 \begin{align}
    \text{SH-CD} =
   \frac
{\sum\limits_{\bk=\pm \text{K}}w(\bk, \ezero,\omega)\,\Omega (\bk,\ezero) }
{\sum\limits_{\bk=\pm \text{K}}w(\bk,\ezero,\omega)\,||\mathbf{d}_{vc}(\bk,\ezero)||^2 } = \abs{\Omega(\pm \text{K})} \, \sigma \, \ezero^2 \, g(\nu) \, f(\omega) + \mathcal{O}(\ezero^4)
\end{align}
with $g(\nu)=\nu / (\Delta ^2 (1-\nu^2))$ and $f(\omega) = 2+\text{Re}\,\eta(\hbar\omega)$ and we write the sum over $\tau=\pm1$ as a sum over $\bk=\pm\text{K}$ and introduce the weight $w(\bk,\ezero,\omega)=|W(\bk)|^2|f_\alpha(\bk)|^2$ defined in equation~\eqref{c8} and ~\eqref{c9}, which is only defined at the $\pm$K points.

\begin{align}
  w(\bk{=}{+}\text{K},\ezero,\omega)=  |W(\tau=+1)|^2|f_\alpha(\tau{=}{+}1)|^2 \\
    w(\bk{=}{-}\text{K},\ezero,\omega)=  |W(\tau=-1)|^2|f_\alpha(\tau{=}{-}1)|^2 
\end{align}

\newpage
\section{S3 Monolayer sample}
\begin{figure}[h!]
    \centering
    \includegraphics[width=\linewidth]{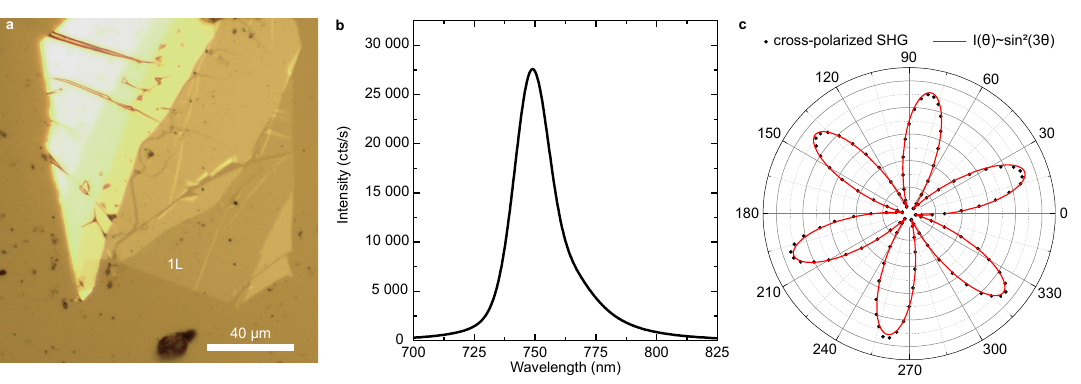}
    \caption{\textbf{Optical image and optical characterization of the monolayer WSe$_2$ sample.} \textbf{a}, Optical microscope image of the exfoliated flake, taken with a 20x objective. The monolayer region is marked with 1L. \textbf{b}, PL spectrum of the monolayer WSe$_2$ sample, with max intensity at \SI{746}{\nm}, corresponding to the A exciton. \textbf{c}, SH polarization dependence of the WSe$_2$ sample in a cross-polarized configuration. The red line shows a $\sin ^2$ fit to the black data points. The FB at $\SI{1500}{\nm}$ is linearly polarized.}
    \label{SI_Fig2}
\end{figure}

\newpage
\section{S4 Polarization of control and fundamental beam}
\begin{figure}[h!]
    \centering
    \includegraphics[width=\linewidth]{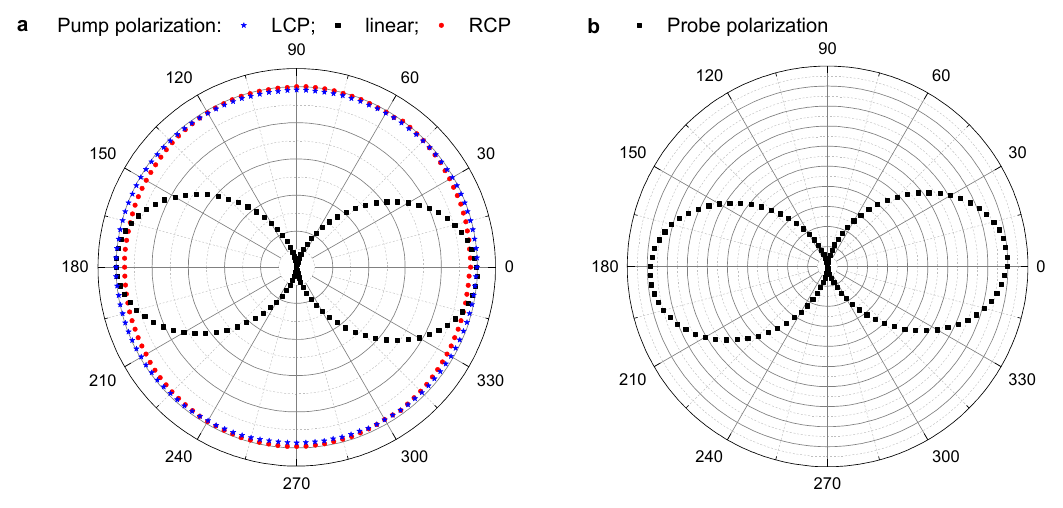}
    \caption{\textbf{Polarization characterization of CB and FB beam} \textbf{a}, Polarization pattern of the CB. LCP (blue) has an ellipticity of 0.98, RCP (red) of 0.99, while linear shows a residual ellipticity of 0.03. \textbf{b}, The polarization pattern of the linearly polarized FB shows a residual ellipticity of 0.02 after reflection at the BS. This small ellipticity underlies the residual SH-CD in Fig.~2a of the main text. All patterns are acquired by rotating a  linear polarizer placed before the focusing objective. The total power is collected on a commercial power head. We define the ellipticity as $\sqrt{I_{\text{min}}/I_{\text{max}}}$.}
    \label{SI_Fig3}
\end{figure}
\newpage

\section{S5 Analytical Model and experimental data}
\begin{figure}[h!]
    \centering
    \includegraphics[width=\linewidth]{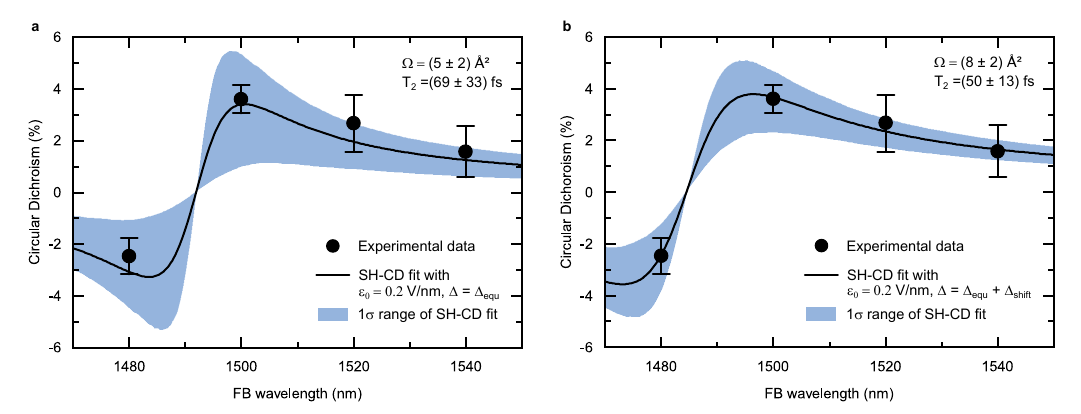}
    \caption{\textbf{SH-CD FB wavelength dependent fit.} Experimental data (circles) and fits (line) of the analytical model in equation~(3) of the main text, with different input parameters for the bandgap $\Delta$. $\Omega$ and $T_2$ are the only fitting parameters. The differences in the optical gap represent the equilibrium bandgap $\Delta_{\text{equ}} = \SI{1.662}{\meV}$, and the valley symmetric bandgap opening induced by the FB~\cite{Klimmer.2026}, $\Delta_{\text{shift}} = \SI{8.2}{\meV}$. The shaded areas represent the $1\sigma$ range of the fit, achieved by Monte Carlo simulation ($N = 1000$) of the parameter uncertainty. The fit algorithm considers the error bars of the experimental data.}
    \label{SI_Fig4}
\end{figure}

\bibliographystyle{naturemag}